\documentstyle[12pt]{article}

\begin{document}

\thispagestyle{empty}

\begin{flushright}
{\footnotesize Int.~J.~Mod.~Phys.~A}
\end{flushright}

\vspace*{5cm}

\begin{center}

{\Large \bf QUATERNION HIGGS\\ AND\\
THE ELECTROWEAK GAUGE GROUP}

\vspace*{2cm}

Stefano DE LEO, Pietro ROTELLI \\

\vspace*{1cm}

{\it Universit\`a  di Lecce, Dipartimento di Fisica\\
I.~N.~F.~N. - Lecce, Italy}

\vspace*{2cm}

{\bf Abstract}

\end{center}
We show that, in quaternion quantum mechanics with a complex geometry, the
minimal four Higgs of the unbroken electroweak the\-ory naturally determine
the quaternion invariance group which corres\-ponds to the Glashow group.
Consequently, we are able to identify the physical significance of the
anomalous Higgs scalar solutions. We introduce and discuss the complex
projection of the Lagrangian densi\-ty.

\pagebreak

\section{Introduction}

\hspace*{5mm} One of the primary objectives of the present authors in
recent years has been to demonstrate the possibility (if not necessity) of
using quaternions in the description of elementary particles, both in
$1^{st}$ and $2^{nd}$ quantization. An essential ingredient in the version of
quaternion quantum mechanics used by the authors is what
Rembeli\'nski\cite{rem} called long ago the adoption of a complex geometry
(complex scalar product). This choice is certainly less ambitious than that
of Adler\cite{adl1,adl2} who advocates the use of a quaternion geometry
and seeks a
completely new quantum mechanics. However, we recall that up to a
decade ago the use of quaternions in $QM$ seemed doomed to failure. The non
commutative nature of quaternions (and hence quaternion wave functions)
made the definition of tensor products ambiguous  and self destructive, e.g.
in general an algebraic product of fermionic wave functions no longer
satisfies the single particle wave equations.

A complex geometry thus seems necessary, if not sufficient, to reproduce
standard $QM$. In fact we have recently shown\cite{del1} that with
the use of generalized quaternions (see Section II) a
translation exists between
{\em even-dimensional} quantum mechanics and our quaternion version. This by
no means concludes the study of this subject. Apart from the eventual
extension beyond standard $QM$ to, for example, the study of intrinsically
quaternion field equations (in the sense in which the Schr\"odinger equation is
intrinsically complex because of the explicit appearance of the imaginary
unit) we have to admit a difference in the bosonic sector (odd-dimensional)
in which additional {\em anomalous} solutions appear. There is also a somewhat
surprising difference in the physical content of Lie group representations
again associated with the odd-dimensional (bosonic) sector, not
withstanding the isomorphism of the corresponding Lie algebras\cite{del2}.

The authors have long been puzzled by the significance of the anomalous
solutions. Although the intial fear of non conservation of momentum has
been overcome\cite{del3}, {\em we have not been able to
identify an anomalous particle before this work}. We had considered the
possibility that with quaternions one
might be able to distinguish between particles
and psuedoparticles. This would be very attractive since in $QM$ the
distinction is by definition. Further\-more, where anomalous solutions do not
occur, such as in the quaternion Dirac equation\cite{rot,mor},
we have as a physical justification
that both parities appear (for particle and antiparticle). However to date
such an identification has not been possible and the results of this paper
lead elsewhere. Indeed we shall argue that to reproduce the
Weinberg\cite{wei}-Salam\cite{sal} model,
or more precisely the Higgs sector, we require anomalous Higgs
solutions of the Klein-Gordon equation and that these are the charged Higgs
that eventually lead to massive $W^{\pm}$ gauge bosons.

Another possible justification for the use of quaternions would be if certain
(correct) choices became {\em natural} with them. Now while we are well
aware that natural\-ness has no rigorous definition and is often synonymous
with habit or some form of analogy, we will argue in just these terms for the
gauge group of the electroweak model.
We shall show that the invariance group of the
Klein-Gordon equation (for a given four-momentum) is $U(1,q)\vert U(1,c)$. The
{\em bar} separates the left-acting unitary quaternion group in one
dimension from the right-acting complex group $U(1,c)$. Here left and
right have nothing to do with helicity. This group substitutes the
Glashow\cite{gla}
group $SU(2) \times U(1)$. We recall that the Lie algebra $u(1,q)$ is
isomorphic to that of $su(2,c)$ as long as one uses antihermitian
generators\cite{del2}. We shall then {\em assume} that this global group is an
invariance group of the Lagrangian density, and this will imply the need of
a complex projection of the dynamic Higgs term. For quite different reasons
a complex projection is needed in the fermionic sector, but this will be
explained elsewhere. Analogy with the standard theory then tells us how we
must proceed for the potential terms.

In the next Section we recall some previous results about the quaternion
Dirac equation. We then discuss the quaternion Kemmer equation and
show that anomalous scalar (and vector) solutions can be avoided if
necessary. We also recall in this Section our rules for translation from
complex to quaternion $QM$ and vice-versa mentioned above. In
Section III we discuss the Higgs particles and derive the above quoted
results. Furthermore we shall obtain a particularly elegant form of
$U(1)_{em}$ and hence the correspon\-ding rule for minimal coupling
which is
a priori ambiguous with quaternions. In Section IV we shall describe the
introduction of the gauge fields by gauging the above group. Our conclusion
are drawn in Section V.

\section{The Dirac and Kemmer Equation}

\hspace*{5mm} We use standard nomenclature for quaternions q,
\begin{equation}
q =  r_{0} + i r_{1} + j r_{2} + k r_{3}
\end{equation}
\begin{center}
$(r_{m}\in {\cal R} (reals) \hspace*{5mm} m=0...3)$
\end{center}
with $i, j, k$ the quaternion imaginary units $i^{2} = j^{2} = k^{2} = -1$
which satisfy
\begin{equation}
ij = - ji = k \; \; \; \; (and \; \; cyclic) \; .
\end{equation}
An alternative (symplectic) decomposition of $q$ is
\begin{equation}
q =  z_{1} + j z_{2}
\end{equation}
\begin{center}
$z_{m}\in {\cal C}(1,i) \hspace*{5mm} m=1,2 \; \; . $
\end{center}
This form implies a choice of one of the imaginary units ($i$ in this case)
which occurs quite naturally in complex wave or field equations such as
Schr\"odinger or Dirac. For us, the unit $i$ will always correspond to the
imaginary unit in standard (complex) $QM$.

We shall now justify the choice of a complex geometry by recalling a
particular derivation of the (irreducible) quaternion Dirac equation. Non
commutativity implies an a priori ambiguity in the form of the Dirac
equation with quaternions. One possibility is
\begin{equation} \label{b}
i {\partial}_{t} \psi = H \psi = ( \vec{\alpha} \cdot \vec{\wp}
+ \beta m) \psi
\end{equation}
where (for covariance arguments) we must define the momentum operator
$\vec{\wp}$ in the standard
way (quaternion hermitian)
\begin{equation}
\vec{\wp} \equiv - i \vec{\partial} \; .
\end{equation}
Unfortunately this choice leads to the non conservation of the norm $N$
\begin{equation}
N = \int  \psi^{+} \psi \; d\vec{x}
\end{equation}
if $H$ is not complex (i.e. if $\vec{\alpha}$ and $\beta$ are not the
standard matrices). Fur\-ther\-more if $H$ is assumed quaternionic then
$\vec{\wp}$
is not even a conserved quantity and so forth. This choice therefore
obliges one to adopt the standard $4$ dimensional complex Dirac matrices. Thus
only the wave function $\psi$ would be quaternion. This use of the standard
$\gamma^{\mu}$ matrices even with quaternions goes back many decades.

An alternative choice which {\em automatically} conserves the
norm is\cite{rot}
\begin{equation}
{\partial}_{t} \psi i = H \psi = ( \vec{\alpha} \cdot \vec{\wp}
+ \beta m) \psi
\end{equation}
this requires for consistency (and covariance)
\begin{equation} \label{a}
\vec{\wp} \equiv - \vec{\partial} \vert i
\end{equation}
where the {\em bar} separates left and right-acting elements. We are thus
lead to define as {\em generalized quaternions} $Q$\cite{del1} the
numerical operators
\begin{equation}
Q = q_{1} + q_{2} \vert i
\end{equation}
\begin{center}
$q_{m}\in {\cal H} \hspace*{5mm} m=1,2$
\end{center}
such that applied to a state vector $\psi$ :
\begin{equation}
Q \psi = q_{1} \psi + q_{2} \psi i
\end{equation}
\begin{center}
$(note \; \; that \; \; q_{1} \equiv  q_{1} \vert 1) \; .$
\end{center}

The choice (\ref{a}) for the momentum operator (in $1^{st}$ quantization)
would not be hermitian unless one introduces the complex scalar product
indicated by the subscript ${\cal C}$
\begin{equation}
(\psi, \phi)_{\cal C} \equiv \frac{1-i\vert i}{2} \, <\psi, \phi>
\end{equation}
where,
\begin{equation}
<\psi, \phi> \equiv \int  \psi^{+} \phi \; d\vec{x}
\end{equation}
is the quaternion scalar product. $H$ may now be quaternion and we are hence
allowed to use the irreducible two dimensional quaternion Dirac matrices which
in turn define a {\em real\/} Dirac algebra. As an aside, we note that
algebraic
theorems which demonstrate that the minimum dimensions of the Dirac
matrices are four, assume the existence of a
{\em complex} Dirac algebra.

Our choice for $\wp^{\mu}$ (${\wp}^0 \equiv {\partial}_{t} \vert i$)
displays the fact that, in order to retain the
standard $QM$ commutation relations, a special imaginary unit must be
selected from the quaternions. Momentum eigenstates will all then be
characterized by standard plane wave functions on the {\em right} of any spinor
or polarization vector (on the left for the adjoint wave functions).

The natural appearance of generalized quaternions has another useful
by-product. While it has long been known that any quaternion can be
represen\-ted by a subset of $2\times2$ complex matrices we are now able to
iden\-tify any two dimensional complex matrix with a generalized quaternion
and viceversa. A particular choice is given by:
\begin{eqnarray}
\begin{array}{ccccccc}
1 & \leftrightarrow  & \left( \begin{array}{cc}
1 & \cdot\\
\cdot & 1 \end{array} \right) & , &
i & \leftrightarrow  & \left( \begin{array}{cc}
i & \cdot\\
\cdot & -i \end{array} \right)\\ \\
j & \leftrightarrow  & \left( \begin{array}{cc}
\cdot & -1\\
1 & \cdot \end{array} \right) & , &
k & \leftrightarrow  & \left( \begin{array}{cc}
\cdot & -i\\
-i & \cdot \end{array} \right)
\end{array}
\end{eqnarray}
and
\begin{eqnarray}
\begin{array}{ccccccc}
1 \vert i & \leftrightarrow & \left( \begin{array}{cc}
i & \cdot\\
\cdot & i \end{array} \right) & , &
i \vert i & \leftrightarrow & \left( \begin{array}{cc}
-1 & \cdot\\
\cdot & 1 \end{array} \right)\\ \\
j \vert i & \leftrightarrow & \left( \begin{array}{cc}
\cdot & -i\\
i & \cdot \end{array} \right) & , &
k \vert i & \leftrightarrow & \left( \begin{array}{cc}
\cdot & 1\\
1 & \cdot \end{array} \right)
\end{array}
\end{eqnarray}
This translation is valid for all operators while for {\em states} we use
the symplectic representation
\begin{eqnarray}
\begin{array}{ccc}
q = z_{1} + j z_{2} & \leftrightarrow  & \left( \begin{array}{c}
z_{1}\\ z_{2} \end{array} \right) \; .
\end{array}
\end{eqnarray}

This still leaves out all odd-dimensional (complex) operators in $QM$
characteristic of the
standard bosonic equations for particles with mass, such as the
Klein-Gordon equation. There is an exception to this last
comment that we wish to investigate further, the Kemmer equation\cite{kem}
\begin{equation}
\beta^{\mu} \partial_{\mu} \phi = m \phi
\end{equation}
(here the $1\vert i$ of the momentum operator has been absorbed into
$\beta^{\mu}$ whose elements must anyway be assumed to be generalized
quaternions). The $\beta^{\mu}$ satisfy the Duffin-Kemmer-Petiau
condition\cite{gen,duf,pet}
\begin{equation} \label{c}
\beta^{\mu} \beta^{\nu} \beta^{\lambda} + \beta^{\lambda} \beta^{\nu}
\beta^{\mu} = - g^{\mu \nu} \beta^{\lambda} - g^{\lambda \nu} \beta^{\mu} .
\end{equation}
This implies that the $\beta^{\mu}$ are not invertible so that this equation
{\em cannot} be written in the Dirac form eq.(\ref{b}). Equation
(\ref{c}) however guarantees that each element of $\psi$ satisfies the
Klein-Gordon equation. The Kemmer equation has spin content 0 and 1 and the
representations for the scalar particle is five dimensional. There exists
however also a trivial one dimensional solution ($\beta^{\mu} \equiv 0$)
which if added to the spin 0 representation yields a six dimensional
representation which can be translated into $3\times3$ generalized quaternions
:
\begin{eqnarray}
\begin{array}{ccccccc}
\beta^{0} & = & \left( \begin{array}{ccc}
\cdot & \cdot & a\\
\cdot & \cdot & \cdot\\
-a & \cdot & \cdot
\end{array} \right) & , &
\beta^{1} & = & j \; \left( \begin{array}{ccc}
\cdot & \cdot & a\\
\cdot & \cdot & \cdot\\
-d & \cdot & \cdot
\end{array} \right)\\ \\
\beta^{2} & = & \left( \begin{array}{ccc}
\cdot & \cdot & \cdot\\
\cdot & \cdot & a\\
\cdot & a & \cdot
\end{array} \right) & , &
\beta^{3} & = & j \; \left( \begin{array}{ccc}
\cdot & \cdot & \cdot\\
\cdot & \cdot & a\\
\cdot & -d & \cdot
\end{array} \right)
\end{array}
\end{eqnarray}
with
\[ a = \frac{1-i\vert i}{2} \; \; \; , \; \; \;
d = \frac{1+i\vert i}{2} \; \; . \]

Now before proceeding we must digress to describe the so called anoma\-lous
solutions, and in particular those of the Klein-Gordon equation. In $QM$ this
equation
\begin{equation}
( \partial^{\mu} \partial_{\mu} + m^{2} ) \phi = 0
\end{equation}
has two solutions (positive and negative energy)
\begin{equation} \label{d}
\phi = e^{-ipx}
\end {equation}
\begin{center}
$p_{0} = \pm \sqrt{{\vec{p}}^{2}+m^{2}} \; .$
\end{center}
With quaternions and with a complex geometry the number of solutions
doubles, in addition to eq.(\ref{d}) we have the complex-orthogonal
solutions
\begin{equation}
\phi = j e^{-ipx}
\end {equation}
\begin{center}
$p_{0} = \pm \sqrt{{\vec{p}}^{2}+m^{2}} \; .$
\end{center}
These are the {\em anomalous} or pure quaternion solutions.

This doubling of solutions does {\em not} occur for our Dirac
equation. This is because the doubling of solutions is compensated there by
the reduced number of spinor components. The question is what happens in the
Kemmer equation? Direct analogy with the Dirac equation is not possible
because the number of solutions no longer correspond to the number of
components of $\psi$. However we can begin with our Kemmer equation, find
the explicit (non trivial) solutions and simply count them or express them
in derivative terms (possible for Kemmer but not for Dirac) so as to obtain
the $2^{nd}$ order equivalent equation.

In fact the solutions to our Kemmer equation are only two,
\begin{eqnarray}
\psi & = & \left( \begin{array}{c}
- \frac{ip_{0}+kp_{x}}{m}\\
\frac{ip_{y}-kp_{z}}{m}\\
1 \end{array} \right) \; e^{-ipx}
\end{eqnarray}
\begin{center}
$p_{0} = \pm \sqrt{{\vec{p}}^{2}+m^{2}} \; .$
\end{center}
This can be rewritten in terms of the $\phi$ in equation (\ref{d})
\begin{eqnarray}
\psi & = & \left( \begin{array}{c}
\frac{(\partial_{t} + j \partial_{x})\phi}{m}\\
\frac{(\partial_{y} + j \partial_{z})\phi}{m}\\
\phi \end{array} \right) \; \; .
\end{eqnarray}
{}From which we derive the necessary and sufficient equation for the scalar
field $\phi$
\begin{equation} \label{e}
\frac{1-i\vert i}{2}( \partial^{\mu} \partial_{\mu} + m^{2} ) \phi = 0 \; .
\end{equation}
This is what we shall call the {\em modified Klein-Gordon equation}. It
does not have anomalous solutions because the projection operator
$\frac{1-i\vert i}{2}$ kills all $j, k$ terms.

There also exists the alternative modified Klein-Gordon equation
\begin{equation} \label{f}
\frac{1+i\vert i}{2}( \partial^{\mu} \partial_{\mu} + m^{2} ) \phi = 0 \; .
\end{equation}
which kills the complex solutions. Note that equations (\ref{e}) and (\ref{f})
are related by a ``quaternion similarity'' transformation
\[ \frac{1-i\vert i}{2} \; \rightarrow \; -j(\frac{1-i\vert i}{2})j =
\frac{1+i\vert i}{2} \]
and
\begin{equation}
\phi \; \rightarrow \; -j \phi \; .
\end{equation}
All of this tells us that we may readily eliminate the anomalous solutions
by invoking the {\em modified bosonic equations}. The correct equation and
the corresponding Lagrangian is thus in practice determined only when the
num\-ber of particles in the theory is fixed. For the Higgs of the next
Section we shall use the standard Klein-Gordon equation which contains four
particles (parity apart).

\section{The Higgs Sector}

\hspace*{5 mm} We know that before spontaneous symmetry breaking the
minimal num\-ber of higgs is four ${\cal H}^{0}, {\cal H}^{+},
\overline{{\cal H}^{0}}, {\cal H}^{-}$. We therefore adopt as a consequence
of the count of states of the previous Section a free Higgs Lagrangian
which yields the Klein-Gordon equation
\begin{equation}
{\cal L}_{free} = \partial_{\mu} \phi^{+} \partial^{\mu} \phi
\end{equation}
where $\phi$ is a massless quaternion field. The field equation
\begin{equation}
\partial_{\mu} \partial^{\mu} \phi = 0
\end{equation}
is obviously invariant under the global group $U(1,q) \vert U(1,c)$
\begin{equation}
\phi \; \rightarrow \; e^{i\alpha +j\beta +k\gamma} \phi e^{-i\delta}
\end{equation}
\begin{center}
$\alpha , \beta , \gamma , \delta$ real parameters.
\end{center}
The limitation of the right-acting group to $U(1,c)$ instead of $U(1,q)$
follows from the implicit additional requirement that the complex plane-wave
struc\-tu\-re be conserved. In $1^{st}$ quantization this would correspond
to maintaining the given momentum. In $2^{nd}$ quantization to the desire of
not assigning a creation or annhilation operator with the incorrect
plane wave structure which would then violate the corresponding Heisenberg
equation\cite{del3} (or yield negative energies).

In order to impose this maximal group invariance of the field
equation upon the free Lagrangian we must {\em assume} that the Lagrangian is
defined as a complex projection (in addition to the hermitian nature of
${\cal L}$ which however involves the creation and annihilation operators).

Thus in fact we can define the Lagrangian density ${\cal L}$ as:
\begin{equation} \label{g}
{\cal L}_{free} = \frac{1-i\vert i}{2} \;
(\partial_{\mu} \phi^{+} \partial^{\mu} \phi) \equiv
(\partial_{\mu} \phi^{+} \partial^{\mu} \phi)_{\cal C}
\end{equation}
This complex projection is automatic for spin $\frac{1}{2}$ fields in order
to reproduce the
standard form of the Dirac equation from the variational principle. In fact
$\psi$ and $\psi i$ must be varied independently in analogy with $\psi^{+}$
and $\psi$, but we shall not enter into detail here.

Any complex projection under extreme right or left
multiplication by a complex number behaves as follows,
\begin{equation}
(z {\cal L} z')_{\cal C} = z ({\cal L})_{\cal C} z' =
z z' ({\cal L})_{\cal C} \; .
\end{equation}
Thus if $zz'=1$ we have invariance. When the transformation is attributed
to the $\phi$ field in eq.(\ref{g}), this implies that $z'=z^{*}$ and hence
\begin{equation}
z \in U(1,c) \; \; .
\end{equation}
It is obvious that if this complex projection is generalized to all terms
in ${\cal L}$ the standard Higgs Lagrangian is an invariant under
$U(1,q)\vert U(1,c)$, since $(\phi^{+} \phi)_{\cal C}$ is\footnote{The quartic
term of the Higgs Lagrangian will also be assumed in the more restrictive form
of $\vert \lambda \vert {(\phi^{+} \phi)_{\cal C}}^{2}$, so
that the plane-wave factors of all $\phi$, $\phi^{+}$ fields may be
factorized as in normal $QM$.}. Hence at this level
our global invariance group is isomorphic with the Glashow group
$SU(2,c) \times U(1,c)$. We must however remember that the group
representations are not totally isomorphic.

In the classical field treatment of spontaneous symmetry breaking we want
the field ${\cal H}^{0}$ to develop a constant {\em real} vacuum
expectation value. This fixes the neutral Higgs to be purely complex fields
(the anomalous fields have no real part). This in turn fixes the
$U(1)_{em}$ gauge which will survive spontaneous symmetry breaking.
Indeed
the requirement of invariance of the neutral Higgs under $U(1)_{em}$
can
only be achieved if ${\cal H}^{0}$ is a complex field in accordance with
the above argument, and
\begin{equation}
U(1)_{em} = e^{ig\alpha}\vert e^{-ig'\delta} \label{p}
\end{equation}
with
\begin{equation}
g\alpha = g'\delta \; ,
\end{equation}
so that the phases cancel after commuting with the Higgs field. The
different signs of the arguments in eq.(\ref{p}) is nothing other than a
convention of the authors. Here we
have explicitly used the (real) coupling strengths $g, \; g'$
characteristic of the Glashow group. We observe the analogy of the above
result with the standard theory, where however one must assume the weak
isospin, weak hyperchar\-ge and electric charge relationship. Here we
appear to have no freedom of choice.

We have a certain number of observations to make,
\begin{enumerate}
\item The above result fixes the mode of minimal coupling (see below).
\item Since under $U(1)_{em}$ the complex Higgs is neutral the anomalous
Higgs (pure $j, k$) are necessarily charged.
\item Had we imposed by fiat that the $U(1)_{em}$ be either the (weak
hyperchar\-ge) right-acting $U(1,c)$ or the left-acting $U(1,c)$ subgroup of
$U(1,q)$, we would have modified the sense of minimal coupling and imposed
a common electric charge on all the Higgs fields. If
$g\alpha \neq g'\delta$ we would also have had four charged Higgs fields
but with different charges.
\end{enumerate}

We return for some further comments upon the complex projection of the
Lagrangian density. We already noted that this condition is obligatory in
the Dirac sector, and therefore it is natural to assume it a property of
the full Lagrangian. One may object that in classical field theory the
Lagrangian is anyway real, so that a complex projection is irrelevant.
However for quantum fields the reality of ${\cal L}$ is substituted by the
hermiticity of ${\cal L}$, so that ${\cal L}$ is {\em not} in general real.
Furthermore, even for classical field theory the reality of ${\cal L}$ does
not in general extend to the variations $\delta {\cal L}$ which may be
complex (and for us even quaternion). Thus it is for these variations that
the complex projection plays a non trivial role.

We conclude this Section by explicitly writing the Higgs part of our
electroweak Lagrangian density
\begin{equation}
{\cal L}^{\cal H} = (\partial_{\mu} \phi^{+} \partial^{\mu} \phi)_{\cal C}
- \mu^{2} \; (\phi^{+} \phi)_{\cal C}
- \vert \lambda \vert \; {(\phi^{+} \phi)_{\cal C}}^{2} \; .
\end{equation}
Note that the quartic potential term is a product of complex projections
and not merely  the complex projection of a product (footnote 1).

\section{Local Group Invariance and minimal coupling}

\hspace*{5 mm} The contents of this Section follows faithfully the standard
procedure, so that we only sketch the various steps. We wish to impose a
local gauge invariance (parameters
$\vec{\theta} \equiv (\alpha, \beta, \gamma)$ and $\delta$
with $x^{\mu}$ dependence). In order to compensate the derivative terms
that then appear in the Lagrangian we introduce four hermitian vector
fields by the following substitution:
\begin{equation}
\partial^{\mu} \rightarrow \partial^{\mu} + \frac{g}{2} \vec{Q} \cdot
{\vec{W}}^{\mu} - \frac{g'}{2} B^{\mu} \vert i
\end{equation}
where $\vec{Q}\equiv (i, j, k)$ are the quaternion imaginary units. The
gauge fields have the well known but peculiar gauge transformation
properties. To find them we impose that
\begin{equation}
({\cal{D}}_{\mu} \phi)' = U ({\cal{D}}_{\mu} \phi)V
\end{equation}
where $U$ and $V$ characterize the transformation of the scalar field
$\phi$
\[ \phi(x) \rightarrow exp(\frac{g}{2} \vec{Q} \cdot \vec{\theta}(x)) \;
\phi(x) \; exp(-i\frac{g'}{2} \delta(x)) = U \phi V \]
and $\cal{D}_{\mu}$ represents the covariant derivative
\[ \partial_{\mu} \rightarrow {\cal{D}}_{\mu} = \partial_{\mu} +
{\tilde{W}}_{\mu} + {\tilde{B}}_{\mu} \]
with
\[ {\tilde{W}}_{\mu} = \frac{g}{2} \vec{Q} \cdot
{\vec{W}}^{\mu} \]
and
\[ {\tilde{B}}_{\mu} = - \frac{g'}{2} B^{\mu} \vert i \; \; .\]
Therefore we have
\begin{equation} \label{m}
({\cal{D}}_{\mu} \phi)' = (\partial_{\mu} U)U^{-1} \phi' +
U(\partial_{\mu}\phi)V
+ \phi' V^{-1}(\partial_{\mu} V) +  {\tilde{W}}'_{\mu} \phi' +
{\tilde{B}}'_{\mu} \phi'
\end{equation}
\begin{equation} \label{n}
 U ({\cal{D}}_{\mu} \phi)V = U(\partial_{\mu}\phi)V +
U{\tilde{W}}_{\mu}U^{-1} \phi' + U{\tilde{B}}_{\mu}U^{-1} \phi' \; .
\end{equation}
By confronting the eq.(\ref{m}, \ref{n}) and noting that $U$ commutes with
${\tilde{B}}_{\mu}$ we find
\begin{equation}
{\tilde{W}}'_{\mu} = U{\tilde{W}}_{\mu}U^{-1} - (\partial_{\mu} U)U^{-1}
\end{equation}
\begin{equation}
{\tilde{B}}'_{\mu} = {\tilde{B}}_{\mu} -
1\vert V^{-1}(\partial_{\mu} V) \; .
\end{equation}
The infinitesimal transformation for the gauge fields are:
\begin{equation}
\vec{Q} \cdot \vec{W}^{\mu} \rightarrow \vec{Q} \cdot \vec{W}^{\mu} -
\vec{Q} \cdot \vec{\theta} +
\frac{g}{2} [ \vec{Q} \cdot \partial^{\mu} \vec{\theta} ,
\vec{Q} \cdot {\vec{W}}^{\mu}]
\end{equation}
\begin{center}
(${\vec{W}}^{\mu} \rightarrow {\vec{W}}^{\mu} - \partial^{\mu} \vec{\theta} +
g \vec{\theta} \wedge  {\vec{W}}^{\mu}$)
\end{center}
and
\begin{equation}
B^{\mu} \rightarrow B^{\mu} - \partial^{\mu} \delta
\end{equation}

Since we have already identified the electro\-magnetic gauge group we can
already
anticipate the residual gauge invariance in terms of the electro\-magnetic
field $A^{\mu}$. By remembering that we can
write\footnote{With our convention  $W_{1}^{\mu}$ (and not  $W_{3}^{\mu}$
as in the standard model) is the neutral member of the {\em weak isospin}
triplet.} $W_{1}^{\mu}$ and $B^{\mu}$ as a linear
combination of $A^{\mu}$ and $Z^{\mu}$
\begin{eqnarray}
\begin{array}{ccc}
B^{\mu} & = & cos\theta_{W} A^{\mu} - sin\theta_{W} Z^{\mu}\\ \\
W_{1}^{\mu} & =  & sin\theta_{W} A^{\mu} + cos\theta_{W} Z^{\mu}
\end{array}
\end{eqnarray}
we have
\begin{equation}
\partial^{\mu} \rightarrow \partial^{\mu} + \frac{e}{2} A^{\mu}
(i-1\vert i)
\end{equation}
\begin{center}
($e$ electric charge)
\end{center}
which can be written in terms of the quaternion projection operator
$(1+i\vert i)$ as
\begin{equation}
\partial^{\mu} \vert i \rightarrow \partial^{\mu} \vert i +
\frac{e}{2} A^{\mu} (1+i\vert i)
\end{equation}
\begin{center}
with $e=\frac{gg'}{\sqrt{g'^{2} + g^{2}}}=gsin\theta_{W}=g'cos\theta_{W}$
\end{center}
but with the understanding that the projection
operator acts upon the scalar field $\phi$,
and guarantees the charge neutrality (invariance) of the pure complex fields.
For simplicity it is convenient to think of the gauge fields as classical
real fields. Thus their position within the Lagrangian density is
irrelevant. In $2^{nd}$ quantization the situation is somewhat more
complicated. The gauge fields are hermitian operator that act upon the kets
(essentially the vacuum) and are {\em bared} operators with the plane wave
structures as right acting factors. In this way standard energy-momentum
conservation occurs without the complication of not commutativity.
Henceforth, unless slated otherwise, we therefore treat the gauge fields as
real classical fields. It is non the less interesting to note
that if the $\vec{Q}$ factors are absorbed within the definition of the
gauge fields then $W_{1}^{\mu}$ and $B^{\mu}$ become complex (indeed pure
imaginary in this classical limit) while $W_{2}^{\mu}$ and $W_{3}^{\mu}$
are both pure quaternion $(j, k)$ or anomalous in our terminology.

We also recall for completeness here the form of the gauge kinetic
terms
\begin{equation}
{\cal L}^{\cal B} = -\frac{1}{4} B_{\mu\nu}B^{\mu\nu} -
\frac{1}{4} {W}_{\mu\nu}^{+} \cdot {W}^{\mu\nu}
\end{equation}
where
\begin{equation}
B^{\mu\nu} = \partial^{\mu} B^{\nu} - \partial^{\nu} B^{\mu}
\end{equation}
and
\begin{equation}
{W}^{\mu\nu} = \vec{Q} \cdot (\partial^{\mu} \vec{W}^{\nu} -
\partial^{\nu} \vec{W}^{\mu}) +
\frac{g}{2} [ \vec{Q} \cdot {\vec{W}}^{\mu} ,
\vec{Q} \cdot {\vec{W}}^{\nu}] \; .
\end{equation}

\section{Conclusions}

\hspace*{5 mm} In this work we have studied the Higgs sector of the
Electroweak theory from the point of view of quaternion quantum mechanics
($QQM$) with a complex geometry. The Higgs fields are assumed to be
four and this coincides with the number of solutions (counting both positive
and negative energies - particles and antiparticles) of the standard
Klein-Gordon
equation {\em within} $QQM$. We have also shown that the quaternion
global invariance group of the one-component Klein-Gordon equation is
$U(1,q) \vert U(1,c)$ isomorphic at the Lie algebra level with the Glashow
group. The right acting phase transformation is limited to the complex
numbers because it must not modify the 4-momentum of the specific solution
considered.

The hypothesis that this group be the invariance group for the Lagrangian
density, then imposes an overall complex projection of the Lagrangian
den\-sity.
This result is {\em consistent} with the need of a complex projection
for the Dirac Lagrangian density in order to obtain the Dirac field
equation. We have pointed out that the reason that a complex projection is
non trivial is because it automatically kills the $j$-$k$ quaternion
variations in $\delta {\cal L}$ which naturally occur when fields and their
adjoint are varied independently.

As an aside we have shown that there exist {\em modified Klein-Gordon
equa\-tions}
(the same applies to Maxwell etc.) with only half of the solutions of the
standard equations. Thus anomalous solutions can always be eliminated
if so desired. Although this result seems obvious a posteriori it was
derived from a study of the quaternion Kemmer equation and we have
sketched the essential steps in Section II. Thus the use of the standard
Klein-Gordon equation is an assumption in $QQM$ with physical consequence
(e.g. the invariance group) and not obligatory for scalars as in complex
quantum mechanics.
It is however to be emphasized that once the desired field
equation has been chosen the invariance group is fixed, unlike the normal
situation, in which there is no relationship between the Klein-Gordon
equation and the multiplicity structure of Higgs fields under
$SU(2) \times U(1)$.

Spontaneous symmetry breaking of the neutral Higgs field then determi\-nes
the resultant form of the residual $U(1)_{em}$. Specifically it is the
complex subgroup of $U(1,q) \vert U(1,c)$
with rotation angles equal in magnitude but
opposite in signs. As a consequence the two other Higgs fields are
anomalous and charged. The significance of anomalous Higgs fields is thus
connected with their electric charge. {\em This is the first time that a
physical property has been associated to pure quaternion fields}.

Our justification of a complex projected Lagrangian density in the case of
the Higgs sector becomes a derivation within the Fermion sector which we
have only outlined in this work. The assumption that all Lagrangian
densities be complex projected then implies that all symmetry groups will
necessarily have a $G\vert U(1,c)$ structure. This is particularly
significant for grand unified theories\cite{lag}. We note that recently much
attention has been paid to the complex group
$SU(5) \times U(1)$\cite{zic}. Within $QQM$ we suggest that
$SU(3,Q)\vert U(1,c) \sim SU(6) \times U(1)$ be a natural candidate for
grand unification.

We conclude by recalling the main point of this work. The need of four
Higgs fields {\em suggests} the use of the standard Klein-Gordon equation.
This equation is invariant under the group $U(1,q)\vert U(1,c)$. The
alternative choice of two modified Klein-Gordon equations would be
invariant only under $U(1,c)\vert U(1,c)$ which beyond being purely complex
in contradiction with the spirit of the use of quaternions and quaternions
groups would not give rise to the charged intermediate vectors bosons. We
the\-re\-fo\-re claim that the natural group of any quaternion
Lagrangian is of the form $G\vert U(1,c)$. The simplest (lowest
dimensional) unitary group being $G=U(1,q)\sim SU(2,c)$. In this sense the
Glashow gauge group appears {\em naturally} as the choice of the minimal
quaternion unitary group for $G$.

\end{document}